\documentclass[11pt,a4paper]{article}
\usepackage{bm}

\newcommand\ie{\textit{i.e.}}

\newcommand\cf{\textit{cf.}}

\newcommand{\half}{\frac{1}{2}}

\newcommand{\be}{\begin{equation}}
\newcommand{\ee}{\end{equation}}
\newcommand{\bea}{\begin{eqnarray}}
\newcommand{\eea}{\end{eqnarray}}
\newcommand{\eqref}[1]{(\ref{#1})}

\newcommand{\dt}{\delta t}
\newcommand{\E}{E_0}

\begin{document}
\begin{titlepage}
\begin{flushright}
%{\tt hep-ph/yymmnn}
{\tt SHEP 11-31}
\end{flushright}

\begin{center}
{\huge \bf  Off-shell OPERA neutrinos}
%\vskip.3cm
%{\huge \bf  and etc} 
\end{center}
\vskip1cm

%\title{Off-shell OPERA neutrinos}
%\author{Tim R. Morris}

\begin{center}
{\bf Tim R. Morris}
\end{center}

%\affiliation{
\begin{center}
{\it School of Physics and Astronomy,  University of Southampton\\
Highfield, Southampton, SO17 1BJ, U.K.}\\
\vspace*{0.3cm}
{\tt  T.R.Morris@soton.ac.uk, Tim.Morris@cern.ch}
\end{center}

%\emailAdd{T.R.Morris@soton.ac.uk}
%\emailAdd{Tim.Morris@cern.ch}

%

\abstract{In the OPERA experiment, superluminal propagation of
neutrinos can occur if one of the neutrino masses is extremely
small. However the effect only has appreciable amplitude at
energies of order this mass and thus has negligible overlap with
the multi-GeV scale of the experiment.}

%\arxivnumber{1110.????}
%\dedicated{preprint SHEP 11-31}

%\notoc
%\begin{document}
%\maketitle
%\flushbottom

\end{titlepage}

\section{Introduction}

Recently the OPERA collaboration reported reported a measurement
of the average time taken for neutrinos ($\nu_\mu$ up to \% level
contamination) created at CERN (CN) to arrive at the Gran Sasso
Laboratory (GS) compared to the time taken travelling at the
speed of light in vacuo ($c$). They found an early arrival time
of approximately $\dt = 60 $ ns, which corresponds, at a
significance of $6.0\sigma$, to faster-than-light travel with
speed $(v-c)/c = 2.48 \times 10^{-5}$ \cite{OPERA}.

Since their announcement, a large number of papers
have
written that variously seek to explain it with
or without new physics,
question the experimental
setup,
or point out difficulties with new
physics explanations (a selection is \cite{some,Glashow,offshell,Naumov,me}).

Here, we work entirely within the framework of standard
relativistic quantum field theory and within the Standard Model
(suitable modified to allow for neutrino masses and mixing matrix).
We note that effective superluminal propagation of $\nu_\mu$ (both
for individual events and, as we will later see, on average), is
possible if the mass $m$ of one of the mass-eigenstates is
extremely small, so small in fact that space-like propagation can
take place with appreciable probability even over the $L=730$ km
distance between CN and GS.

To see that this is possible in principle, consider a neutrino of
such a mass created at CN at time $x^0=0$ and position $x=0$, and
arriving at GS with space-time coordinates
$(y^0,y)=(L-\dt,L)$.{\footnote{We can neglect the two transverse
spatial coordinates. From hereon we work in units with $\hbar=c=1$,
translating back as necessary.}} To describe this we should use the
Feynman propagator for spinors. However the neutrino beam has
average energy $\langle E\rangle = 17$ GeV \cite{OPERA}. The
neutrinos are created left handed (by weak decay of mesons in the
decay tunnel) and being ultra-relativistic, will stay that way to
very good approximation throughout their flight. Therefore we need
only the left handed component, which effectively reduces the
propagator to that of a scalar particle:
\begin{equation}
\label{Fprop}
S(y-x) = \lim_{\epsilon\to 0} \int\!\! {d^4p\over (2\pi)^4}\,
\frac{ {\textrm{e}}^{-ip\cdot(y-x)} }{p^2-m^2 + i\epsilon}
\end{equation}
(Here $p^\mu$ is
the four-momentum.) When the interval $s := (x^0-y^0)^2 -
(x-y)^2$ is negative as is measured by OPERA,\, \ie\ is
space-like, then we have that
\begin{equation}
S(y-x) = -\frac{im}{4\pi^2{\sqrt{-s}}} K_1(m\sqrt{-s}),
\end{equation}
where $K_1$ is a modified Bessel function. For large value of its
argument it decays exponentially as:
\begin{equation}
   \label{decay}
    K_1(m\sqrt{-s})\sim
    \sqrt{\frac{\pi m}{2}} {1\over(-s)^{1\over4}}\,\textrm{e}^{\textrm{\normalsize $-m\sqrt{-s}$}}.
\end{equation}
On the other hand for small values, $K_1$ diverges:
\begin{equation}
K_1\sim \frac{2i}{\pi m\sqrt{-s}}.
\end{equation}
We see that if we choose
\begin{equation}
\label{firstestimate}
    m\sim \frac1{\sqrt{-s}} \approx \frac1{\sqrt{2L\dt}} = 3.8\times 10^{-11}
    {\textrm{eV}},
\end{equation}
then a substantial fraction of the neutrinos can propagate
superluminally.

Of course it is not the case here that the neutrinos are really
tachyonic. The effect occurs because the neutrino remains off
shell.\footnote{It follows that the constraints noted in ref. \cite{Glashow} do not apply: kinematically electron-positron bremsstrahlung would here require momentum transfer with constituents of rock, and is just one of many diagrams heavily suppressed by the Fermi constant and the requirement that the neutrino remains off shell.} The Feynman propagator arises from the Lorentz
covariant time-ordered expectation of
the $\nu_\mu$ and ${\bar\nu}_\mu$ fields
\bea
\label{nuprop}
S(y-x) &=& \langle0|T\nu(x){\bar\nu}(y)|0\rangle\nonumber\\
&=&  \theta(x^0-y^0)\langle0|\nu(x){\bar\nu}(y)|0\rangle -\theta(y^0-x^0)  \langle0|{\bar\nu}(y)\nu(x)|0\rangle.
\eea
The propagator arises as an intermediate step in  a chain of reactions, for example being created at CN through the decay $\pi^+\to\mu^+\nu_\mu$, and absorbed at GS through $\nu_\mu n\to \mu^- p$.
 
We can interpret this as the $\nu_\mu$ being created with energy-momentum $p^\mu=\langle E\rangle(1,v)$.  The faster-than-light propagation occurs because the field operators are not localised but have a spread of order the Compton wavelength $1/m$, which in our case is large enough to stretch from CN to GS. 

Since we know that $y^0-x^0>0$, \ie\ that the neutrino arrives in GS after being created at CN, the first term currently plays no r\^ole. 
The potential problems with causality and Lorentz covariance arise when we view the propagation in a frame moving at speed $v_F$ sufficiently fast that the event at GS happens before the one at CN: $y'^0 = \gamma_F(y^0-v_Fy)=\gamma_F y^0(1-vv_F)<0$ (where the dilation factor $\gamma_F=1/\sqrt{1-v^2_F}$). Since the Feynman propagator is Lorentz covariant, it still takes the same form, but in this frame it is the first term in \eqref{nuprop} that now operates. It describes a superluminal antineutrino ${\bar\nu}_\mu$ travelling from GS to CN. Energy and momentum conservation are preserved in the chain of reactions, and moreover we see that the neutrino energy has also reversed sign: $E'=\gamma_F\E(1-vv_F)<0$. Therefore we have a consistent interpretation in this new frame: a ${\bar\nu}_\mu$ is created at GS with energy $-E'$ and momentum pointing from GS to CN, through the process $n\to p\mu^-{\bar\nu}_\mu$, and absorbed at CN through the process ${\bar\nu}_\mu\pi^+\to\mu^+$. The kinematics of these processes are allowed because the anti-neutrino's 4-momentum is space-like. 

This is of course just the standard argument due to Feynman with small adaptations to this novel context, and demonstrates that the Feynman propagator allows an interpretation consistent with Lorentz covariance. However, there still remains an issue of causality. Indeed, in the new frame it seems the neutron at GS has somehow to know to decay beforehand so that the antineutrino can later be absorbed by the pion produced at CN. 

Once we have understood the process in more detail quantum mechanically, we can see how this issue with causality is resolved. We will return to this in the conclusions.

While this paper was being prepared, ref. \cite{offshell} appeared,
where similar ideas are proposed as an explanation for the OPERA
measurement. In fact, if we adapt this observation to the setup in
the OPERA experiment, we can see that the effect vanishes
or at best is far too small to explain the measurement.

The reason our setup is not yet the pertinent one is that it is not
true that the neutrinos are created at an exact time. By
Heisenberg's uncertainty principle, such a particle would have an
infinite spread in energy. In sec. 2 we argue that the spread
in energy for each neutrino's wave packet is $1/\tau\sim0.2$ eV
about its central value $E_0$. In the ensemble of neutrinos that
make up the neutrino bunches, these central values are spread over
a wide range \cite{OPERA,talk} but it is clear nevertheless that
$E_0$ is set at a much larger scale (GeV to multi-GeV).

Given that each neutrino's energy is thus sharply peaked we can already expect that the effect must be heavily suppressed. From \eqref{decay}, the evanescent part of the
propagator appears dominated by energies of order
\eqref{firstestimate} -- indeed we will see that it is a manifestation of quantum
mechanical tunnelling, requiring energy to be less than $m$; 
this mismatch with the energies in the wavepacket $\sim
E_0\pm1/\tau$, ensures that any remaining effect is consigned to
any small tail in the wavepacket that reaches down to these small
energies. For example, if we assume that the wavepacket is a
Gaussian of width $1/\tau$ then this supplies a suppression factor
\be
\label{dead}
\sim\exp-\tau^2\E^2\sim\exp-\tau^2\langle E\rangle^2\sim {1\over10^{10^{10}}}.
\ee 
Alternatively, a wave packet with a lower cutoff $>m$ on the neutrino energy would eliminate the effect entirely.
Furthermore, the space-like neutrino then
has to be detected at Gran Sasso, which requires overlap of
\eqref{decay} with the typical wavefunction for a detected
neutrino.\footnote{more accurately the amplitude constructed from
the detected interaction products} Such a wavefunction has again an
order GeV central energy and spread $\sim1/\tau$. (At the most
optimistic we can note that a lower bound is set by the lowest
threshold energy for detection of $\nu_\mu$ via charge current
interactions in the Gran Sasso Laboratory. This is for the process
$\nu_\mu n\to \mu^- p$, thus $E_0 > m_\mu+{\half m_\mu^2/m_n}=112$
MeV.) This results in multiplying by another suppression factor
$\sim\exp-\tau^2\E^2$ from the tail of its distribution.

One might try to resuscitate the proposal by bringing $1/\tau$ and $E_0$ closer to each other. The relevant $E_0$  could be as low as the threshold energy of $112$ MeV and perhaps we have missed something in our estimate of the energy spread $1/\tau$ in the neutrino wave packet and the true spread is far larger.
If we for the moment take this optimistic view, we note that we still cannot fit the data. The dominance
of the very low energy scale is not seen at OPERA: they
repeated the analysis concentrating on only those $\nu_\mu$ charged
current events occurring in the OPERA target (where reliable
energies could be measured). They split this sample into bins of
nearly equal statistics, taking events with energy higher or lower
than 20 GeV. With a significance of greater than 3$\sigma$, they
still see a superluminal velocity in the higher energy sample of
the same magnitude (within errors) \cite{OPERA}.

Another way one might try nevertheless to use this effect to explain the OPERA measurement is to boost \eqref{Fprop}. In other words, we note that the neutrinos are neither created at an exact time nor at an exact location; in reality we need to integrate over position space terms that supply the neutrino with the appropriate ultra relativistic  momentum. One could then hope that the behaviour \eqref{decay} would be the correct one for directions transverse to the neutrino's momentum, corresponding to small deficits in the energy required by the ultra relativistic on-shell energy-momentum relation. The amplitude would effectively take the form of the kilometres-wide wave function assumed  in ref. \cite{Naumov}, where the OPERA result is then explained without any adjustable parameters by the off-centre detection of these wave packets. However we will see that this set-up does not result in such transverse evanescence. 
%of the Feynman propagator.

%In the remainder of the paper we present the calculation. This will
%confirm the intuitive arguments above and determine the effect in general
% so that it can be adapted to other scenarios where
%off-shell particles of almost certain energy might play a r\^ole.

Although neutrino species oscillate into one another, we can ignore
this effect in the present paper. The known mass-squared
differences \cite{PDGrev} show that only one mass eigenstate can be
as light as \eqref{firstestimate}. Therefore the effect we are
looking for is entirely due to evanescence in this mass eigenstate.
The only effect of mixing is the multiplicative inclusion of mixing
angles at the beginning where the CN $\nu_\mu$ converts to this
state, and at the end where it reverts to $\nu_\mu$, as detected in
GS.

The structure of the paper is as follows. In the next section (sec.
2) we estimate the initial spread in energy of the neutrino
wavefunction at CN and the effective further decoherence due to its
detection at GS.  In sec. 3, we construct the initial wavefunction
and the amplitude for neutrinos as seen at the Gran Sasso
Laboratory, and draw out the piece relevant for this effect, confirming and extending the arguments given above. Finally, in sec. 4, we present our conclusions and also argue why these results ensure there is no violation of causality.

\section{Estimating coherence}

We have already noted that we cannot assume that the neutrino is
created at a definite time. Likewise however, it is not true that
each neutrino has been created with a definite energy. Such a
particle would be completely delocalised in time, so the question
of when it arrives in Gran Sasso would become meaningless. We
therefore have no choice but to take into account the shape of its
wavepacket, at least in its gross details, since the results will
depend on this.

It should be clear that here we are not discussing the energy spectrum of the ensemble of neutrinos in the beam, which is very broad, but rather the inherent quantum mechanical uncertainty in the momentum of a given neutrino when it is created at CN and any further inherent uncertainty imposed on it by being measured at GS.

We follow closely the analysis in ref. \cite{me}. We start at CERN
with a proton bunch extracted from the Super Proton Synchotron.
%with the amplitude $\Psi_{CN}(t,0)$ to produce a neutrino in the CNGS decay tunnel at CERN.
The energy uncertainty inherent in an individual proton wavefunction  can certainly can be no smaller than that set by $\Delta t \approx 5$ ns, the smallest time features in the proton bunch \cite{OPERA}. This corresponds to $\Delta E=1/\Delta t\approx1.3\times10^{-7}$ eV.

However, even if the proton beam is coherent at this level, it suffers decoherence on its way to becoming the $\nu_\mu$ beam. Firstly, the protons impact the graphite target, producing the mesons (mostly pions) that will decay to muon neutrinos. Initially these mesons are in a quantum state together with the other products of the collision (including various nuclei), however they then suffer decoherence from thermalisation in a hot target, both directly and also through their quantum mechanical coupling to the decay products. Assuming a target temperature of, say 300$^\circ$C, this limits the energy-momentum resolution to $k_BT\sim0.05$ eV.

Finally the mesons decay in the decay tunnel and here further decoherence takes place, again through coupling to the decay products (in this case the muon). Consider for example the decay of a $\pi^+$ to $\mu^+\nu_\mu$. The resulting quantum state takes the form:
\be
\label{pidecay}
\psi_\pi({\bf q})\int_{ {\bf p}\ phase\ space}\!\!\!\!\!\!\!\! \mathcal{M}({\bf p},{\bf q})\  |\nu_\mu({\bf p})\rangle\,|\mu^+({\bf q}-{\bf p})\rangle,
\ee
where $\psi_\pi({\bf q})$ is the wave function of the erstwhile pion,  ${\bf q}$ and ${\bf p}$ are 3-momenta, and $\mathcal{M}$ stands for the matrix element for the decay. The muon is absorbed by a combination of rock, a Hadron stop and two muon detectors \cite{OPERA,CNGStalkonwebsite}. This allows the experimenters to measure the transverse coordinates of the proton beam spot when it hits the target to a precision of $\sim$ 50 -- 90 $\mu$m \cite{CNGStalkonwebsite}, however it is reasonable to assume that the rock itself localises the muons at the $\mu$m level (similar to emulsion --- see below), even if this is not recorded. This corresponds to a momentum decoherence of order 0.2 eV/c. Through \eqref{pidecay} this decoherence is transferred to the neutrino.

We conclude that the chief limiting factor on the coherence of the
initial neutrino wave packet is through the decay of the mesons and
results in a wave packet with an energy spread of $1/\tau\sim$ 0.2
eV.

For muon neutrinos that interact in the Gran Sasso detector, the
impact spot is discernible in the emulsion at the $\mu$m level
\cite{impactspot}; we can expect similar localisation in the rock
in front of the detector for the external events. Therefore the act
of measurement results in an effective energy spread in the wave
packet of similar size to that in the initial packet.

\section{Tunnelling to Gran Sasso}

We can thus regard our neutrino as being created at position $x=0$
at an uncertain time centred around $t=0$, with energy localised to
$E=E_0$ with an accuracy $1/\tau$. Let us model the shape as a
Gaussian. Then we have for the initial wavefunction:
\be
\label{init}
\Psi(0,t) \propto \exp\left\{-{t^2\over2\tau^2}-iE_0 t\right\}.
\ee
The amplitude to find the neutrino at Gran Sasso at time $t_2=y^0$
is then
\be
\label{intovert}
A(t_2)
\propto\int^\infty_{-\infty}\!\!\!dt\,\,\Psi(0,t)\,S(t_2-t,L),
\ee
where $S$ is the Feynman propagator \eqref{Fprop}. Since we are
dealing with a situation where we know the neutrino arrives at Gran
Sasso, we will choose the normalization factor so that the probability distribution reflects this:
\be
\int^\infty_{-\infty}\!\!\!dt_2\,\,|A|^2(t_2) = 1.
\ee
We can now make a further simplification. We can drop the
transverse momentum integrations in \eqref{Fprop} since these will
only result in $\sim 1/L^2$ losses that are scaled away when we
normalise. Therefore, substituting \eqref{Fprop} and using
\eqref{init}, we have
\bea
\label{intprop}
A(t_2) &\propto&
\int^\infty_{-\infty}\!\!\!dt\,\,\Psi(0,t)\,\int^\infty_{-\infty}\!\!\!dE\,dq\,\,
{\exp\left\{-iE[t_2-t]+iqL\right\}\over E^2-q^2-m^2+i\epsilon}\nonumber\\
 &\propto&
 \int^\infty_{-\infty}\!\!\!dE\,dq\,\,{{\exp\left\{
-iEt_2-\tau^2\Delta E^2/2+iqL\right\}}\over
E^2-q^2-m^2+i\epsilon},
\eea
where $\Delta E = E- E_0$, and we have performed the $t$ integration. Now we do the momentum integration.
Since we know that $L>0$, the $i\epsilon$ prescription tells us to
close the contour over the top. We obtain:
\be
\label{ampE}
A(t_2) = N \int^\infty_{-\infty}\!\!\!\!dE\,\,
\textrm{e}^{-iEt_2-\tau^2\Delta E^2/2}\, {\textrm{e}^{ipL}\over p},
\ee
where $N$ is the normalisation constant, where $p=\sqrt{E^2-m^2}$ has either a vanishing positive imaginary
part (the $i\epsilon$) for $|E|>m$, or $p=i\sqrt{m^2-E^2}$ for
$|E|<m$.

We see that the exponential decay in \eqref{decay} indeed
arises from energies of order $m$, as we claimed. We also confirm that the exponential decay component is suppressed into the far tail of the probability distribution by $\sim\exp-\tau^2 E^2_0$. The amount of evanescent component thus depends crucially on the unknown shape of this tail. If we had chosen a wave packet with a sharp cutoff at some realistic minimum energy, we would eliminate the evanescent contribution completely.

For completeness we note that \eqref{ampE} can be evaluated by the method of steepest descents. The dominant term comes from $E\approx E_0$ giving:
\be
\label{dominant}
A(t_2)|_{dominant}\approx N{\sqrt{2\pi}\over p_0\tau}\exp-\left\{{1\over2\tau^2}\left(t_2-{L\over v_0}\right)^2+i(E_0t_2-p_0L)\right\}.
\ee
(The approximation is valid providing $|t_2-L/v_0|\ll \tau^2 p_0$.)
Here $p_0 =\sqrt{E^2_0-m^2}$ is the central momentum and $v_0 =p_0/E_0$ is the classical velocity. We recognise this as nothing but the expected result of propagating the wave packet to Gran Sasso without dispersion. The evanescent part can also be evaluated. Writing $E=m(1-z)$ for small positive $z$ at the top boundary, the integral can again be evaluated by steepest descents. Adding to this the term from the bottom boundary $E=-m(1-z)$, and similar size pieces from $E=\pm m(1+z)$ (which can again be evaluated by steepest descents), we get for the evanescent part
\be
A(t_2)|_{evanescent} \approx -2iN \sqrt{2\pi\over mt_2}\textrm{e}^{-\tau^2 E^2_0/2} \exp i\left\{{mL^2\over2t_2}-m t_2+{\pi\over4}\right\}.
\ee
This approximation is valid providing $mt_2\gg1$. We see again the damping by the tail of the wave function. We can now carry this through to a computation of the superluminal component of velocity. The evanescent term will provide one, but we do not present the computation since we have seen that it necessarily depends on a vanishingly small unknown quantity.

Finally, we address the question raised in the introduction: whether, after taking into account appropriate spatial dependence of $\Psi$  and spatial integration in \eqref{intovert}, so as to incorporate the fact that the neutrino has  some momentum $\mathbf{\tilde p}_0$ (with associated small uncertainty) which is slightly larger in magnitude than the energetically allowed $p_0 =\sqrt{E_0^2-m^2}$, the resulting amplitude could take the form similar to \eqref{dominant} in the direction of $\mathbf{\tilde p}_0$ but with transverse evanescent tails corresponding to imaginary transverse momenta? Recall that such an amplitude would be very similar to the wave packet envisaged in ref. \cite{Naumov} where results consistent with the OPERA measurement were derived as a result of off-centre measurement of these wave packets.

Clearly in order to investigate this we now need to keep the transverse momentum integrations in \eqref{Fprop}. The modified wave function and spatial integration incorporated in \eqref{intovert} results in a momentum integral in \eqref{intprop} of the form:
\be
J(\mathbf{L}) = \int\!\!d^3q\, \Phi(\mathbf{q}) {\exp\{i\mathbf{q}\cdot\mathbf{L}\}\over E^2-\mathbf{q}^2-m^2+i\epsilon},
\ee
where $\Phi$, strongly peaked about $\mathbf{q}= \mathbf{\tilde p_0}$ incorporates the momentum dependence induced by the initial wave function. 

Now we appeal to the Grimus-Stockinger theorem \cite{Grimus} which states that\footnote{providing $\Phi$ satisfies the reasonable conditions that it is three-times continuously differentiable and first and second derivatives decrease at least as $1/\mathbf{q}^2$ for $|\mathbf{q}|\to\infty$} 
\be
J(\mathbf{L}) = -{2\pi^2\over L}\Phi\left(-p{\mathbf{L}/ L}\right)\exp\{ipL\}+O(L^{-3/2}),
\ee
if $p=\sqrt{E^2-m^2}$ is real, and is $O(1/L^2)$ otherwise. We see that providing $E>m$ we get only the analogous dependence to \eqref{dominant}. Once we integrate over energy in \eqref{intprop}, the mismatch between $p_0$ and ${\tilde p}_0$ will result in suppression arising from the partial overlap of $\exp-\tau^2\Delta E^2/2$ and $\Phi\left(-p{\mathbf{L}/ L}\right)$, however we do not generate evanescent tails as a result of this mismatch. On the other hand, for $E<m$, $J$ decays faster in $L$, which we can associate with the evanescence. We see that even with this `boosted' wave packet, it is still the case that the evanescent behaviour responsible for superluminal propagation, is restricted to the regime $E<m$.

\section{Conclusions}

We have seen that although superluminal propagation is possible over the 730 km from CERN to Gran Sasso if the mass of the lowest neutrino mass eigenstate is so small that it remains off shell, \cf\, \eqref{firstestimate}, the effect does not survive projection on the relevant energy scales, being killed dead or by typically a huge suppression \eqref{dead}. This projection is required because one must integrate over the initial neutrino wave function which carries a very rapidly oscillating exponential, set by the multi-GeV energy of the created neutrino. One could reduce the suppression (\ie\, increase the overlap with energy scale $m$) by concentrating on muon neutrinos with energies only just higher than the threshold energy of $112$ MeV. This is still not enough: one would have to argue that the quantum-mechanical energy uncertainty in the neutrino wave packet was not of order 0.2 eV as we have argued but -- implausibly -- tens or hundreds of MeV. Even if one does this, the effect is still concentrated at energies of order $m$, resulting in the wrong energy dependence.

One could reduce the time for which the neutrino has to remain off shell by considering cases where it propagates on shell to a point near the detector and, due to interaction with the rock, is then kicked off shell for the remaining part of its journey. Note that an off-shell neutrino can be passed from this point to the detector in principle almost instantaneously. In this way we achieve the smallest negative interval $s$ from the instantaneous jump $(0,\dt)$ from $(L-\dt,L-\dt)$ to the detector. In \eqref{firstestimate} we then get the larger mass $m\sim 1/\dt \sim 10^{-8}$ eV \cite{offshell}. This is still far too small to bridge the gap to the energy scales set by the experiment.

%Finally, we return to the issue of violations of causality which we %discussed in the introduction. 
Although the OPERA measurements cannot be explained by assuming very low mass off-shell neutrinos, it is still the case that such neutrinos can propagate superluminally.  How can this be reconciled with causality, in particular why does this not lead to faster-than-light communication? Note that we have shown that these neutrinos must carry energy $E$ less than $m$. Leaving aside practical issues involved in detecting such neutrinos, we note that even in principle, a detector for such neutrinos would have to be restricted to using wavelengths larger than or of order their Compton wavelength $\lambda=1/m$, otherwise the observation itself would disturb the system too much -- for example by pair creating the very neutrinos it was trying to measure. Thus we see that the restriction to $E<m$ for superluminal neutrinos, which comes out from the detailed analysis, is in fact necessary to ensure that they cannot be detected with sufficient spatial resolution to allow faster than light signals.

%In this way, we see that quantum field theory ensures there is  %operationally the effect cannot be used to send signals faster than light.

\section*{Acknowledgments}
The author  thanks the STFC for financial support and Konstantin Kuzmin and Bartolome Alles Salom for useful conversations.


\begin{thebibliography}{99}

%\cite{:2011zb}
\bibitem{OPERA}
  T.~Adam {\it et al.} [ OPERA Collaboration ],
  ``Measurement of the neutrino velocity with the OPERA detector in the CNGS beam,''

  [arXiv:1109.4897 [hep-ex]].
  
\bibitem{some}
D.~Fargion and D.~D'Armiento,
  ``Inconsistence of super-luminal Opera neutrino speed with SN1987A neutrinos
  burst and with flavor neutrino mixing,''
  arXiv:1109.5368 [astro-ph.HE];
  %%CITATION = ARXIV:1109.5368;%%
   G.~F.~Giudice, S.~Sibiryakov, A.~Strumia,
  ``Interpreting OPERA results on superluminal neutrino,''
  [arXiv:1109.5682 [hep-ph]];
V.~K.~Oikonomou,
  ``The 2d Gross-Neveu Model for Pseudovector Fermions and Tachyonic Mass
  Generation,''
  arXiv:1109.6170 [hep-th];
 J.~Alexandre, J.~Ellis, N.~E.~Mavromatos,
  ``On the Possibility of Superluminal Neutrino Propagation,''
  [arXiv:1109.6296 [hep-ph]];
  %%CITATION = ARXIV:1109.6170;%%
P.~Wang, H.~Wu and H.~Yang,
  ``Superluminal neutrinos and domain walls,''
  arXiv:1109.6930 [hep-ph];
  %%CITATION = ARXIV:1109.6930;%%
  M.~A.~Anacleto, F.~A.~Brito and E.~Passos,
  ``Supersonic Velocities in Noncommutative Acoustic Black Holes,''
  arXiv:1109.6298 [hep-th].
  %%CITATION = ARXIV:1109.6298;%%
  G.~Henri,
  ``A simple explanation of OPERA results without strange physics,''
  [arXiv:1110.0239 [hep-ph]];
  W.~Winter,
  ``How large is the fraction of superluminal neutrinos at OPERA?,''
  [arXiv:1110.0424 [hep-ph]];
  M.~V.~Berry, N.~Brunner, S.~Popescu and P.~Shukla,
  ``Can apparent superluminal neutrino speeds be explained as a quantum weak
  measurement?,''
  J.\ Phys.\ A  {\bf 44} (2011) 492001
  [arXiv:1110.2832 [hep-ph]].
  %%CITATION = JPAGB,A44,492001;%%
 F.~Giacosa, P.~Kovacs and S.~Lottini,
  ``Could the OPERA setup send a bit of information faster than light?,''
  arXiv:1110.3642 [hep-ph].
  %%CITATION = ARXIV:1110.3642;%%
J.~Bramante,
  ``Sterile Neutrino Production Through a Matter Effect Enhancement at Long
  Baselines,''
  arXiv:1110.4871 [hep-ph].
  %%CITATION = ARXIV:1110.4871;%%
  
  
  
  %\cite{Cohen:2011hx}
\bibitem{Glashow}
  X.~J.~Bi, P.~F.~Yin, Z.~H.~Yu and Q.~Yuan,
  ``Constraints and tests of the OPERA superluminal neutrinos,''
  arXiv:1109.6667 [hep-ph];
  %%CITATION = ARXIV:1109.6667;%%
A.~G.~Cohen, S.~L.~Glashow,
  ``New Constraints on Neutrino Velocities,''
  [arXiv:1109.6562 [hep-ph]].



%\cite{Ahluwalia:2011rg}
\bibitem{offshell}
  D.~V.~Ahluwalia, S.~P.~Horvath, D.~Schritt,
  ``Probing neutrino masses with neutrino-speed experiments,''
  [arXiv:1110.1162 [hep-ph]].
  
\bibitem{Naumov}
  D.~V.~Naumov and V.~A.~Naumov,
  ``Neutrino Velocity Anomalies: A Resolution without a Revolution,''
  arXiv:1110.0989 [hep-ph].
  %%CITATION = ARXIV:1110.0989;%%
  
%\cite{Morris:2011sn}
\bibitem{me}
  T.~R.~Morris,
  ``Superluminal group velocity through maximal neutrino oscillations,''
  
  [arXiv:1110.2463 [hep-ph]].


\bibitem{talk} Dario Autiero ``Measurement of the the neutrino velocity with the OPERA detector in the CNGS beam,''
http://cdsweb.cern.ch/record/1384486

\bibitem{PDGrev} K. Nakamura \emph{et al.} (Particle Data Group), J. Phys. G\textbf{37}, 075021 (2010) and 2011 partial update for the 2012 edition.

  
\bibitem{CNGStalkonwebsite}
A.~E.~Ball {\it et al.}, ``CNGS muon monitoring: Why do we need both muon detector stations at the start-up of the CNGS neutrino beam?,'' CERN SL Division technical note SL-Note-2002-040 EA EMDS No. 36258 available on the CNGS website http://proj-cngs.web.cern.ch


%\cite{Agafonova:2009kr}
\bibitem{impactspot}
  N.~Agafonova, A.~Anokhina, S.~Aoki, A.~Ariga, T.~Ariga, L.~Arrabito, D.~Autiero, A.~Badertscher {\it et al.},
  ``The Detection of neutrino interactions in the emulsion/lead target of the OPERA experiment,''
  JINST {\bf 4 } (2009)  P06020.
  [arXiv:0903.2973 [hep-ex]].

\bibitem{Grimus}
  W.~Grimus and P.~Stockinger,
  ``Real oscillations of virtual neutrinos,''
  Phys.\ Rev.\  D {\bf 54} (1996) 3414
  [arXiv:hep-ph/9603430].
  %%CITATION = PHRVA,D54,3414;%%

\end{thebibliography}
\end{document}